\documentclass[10pt,a4paper, dvipdf]{article}
\usepackage[hmargin=3cm, vmargin=1.5cm]{geometry}
\usepackage{textcomp}
\usepackage{subfigure} 
\usepackage{xspace}
\usepackage [T1]{fontenc} 
\usepackage{amssymb}
\usepackage{hyperref}
\usepackage[ansinew]{inputenc}
\usepackage{array}
\usepackage[all]{xy}
\usepackage{graphics}
\usepackage{multicol}
\usepackage [sort&compress]{natbib}
\bibpunct{[}{]}{,}{n}{}{;} 
\newcolumntype{C}{>{$}c<{$}}
\newcommand {\EE}{{ \mathbf {E}}}
\newcommand { \s} {{ \mathrm {S}}}

\newcommand {\pr} {{ \mathrm {Prob}}}
\newcommand {\ordinal} {{ \mathrm {th}}}
\newcommand {\PP} {{ \mathrm {P}}}
\newcommand{\br}{\hline}
\newcommand{ \sr}{\hline}
\newcommand{\Tref}{Table \ref}
\newcommand{\Eref}{Equation \ref}
\newcommand{\Fref}{Figure \ref}

\title{\bf{Classical counterexample to Bell's theorem}}
\author{Michel Feldmann} 
\date{}
\begin{document}
 \maketitle

\abstract{
We describe a strictly classical dice game, which emulates the main features of the EPR experiment, including violation of Bell's inequalities. Therefore, the standard interpretation that Bell's theorem provides necessary conditions for `local realism' is disproved. 

} 
%
%
\section{Introduction}
\label{introduction}
In his celebrated theorem on hidden variables, John Bell~\cite{bell, bell1} proved in 1964 that quantum mechanics violates some inequalities. He then concluded that quantum mechanics cannot be reproduced by local hidden variable theories of any kind, in total opposition with Einstein conception~\cite{einstein}. However, this idea was supported by a number of experimental verifications following the famous experiment of Aspect et al.~\cite{aspect} even if `a conclusive experiment falsifying in an absolutely
uncontroversial way local realism is still missing'~\cite{genovese}.  Violation of local realism in quantum mechanics  is now widely accepted by the scientific community~\cite{GHZ}. Nevertheless, several authors remain reluctant to abandon this property~\cite{santos} and some even contest the validity of Bell's theorem~\cite{clearing,mf,kracklauer,hess1,khrennikov, khrennikov2,cameleon,clover1,nieuwenhuizen,adenier}. Of course, this theorem is a perfect mathematical result and only Bell's \emph{hypotheses} are under discussion. The irrefutable technical criticism came in 1988 by E.~T.~Jaynes~\cite{clearing} in an opening talk at a Cambridge conference on Bayesian Methods.  Jaynes pointed out that the fundamentally correct relation according to probability theory should make use of conditional probabilities while, by contrast, Bell postulates the existence of an absolute probability space, irrespective of the settings. This flaw in the assumptions of Bell's theorem was also considered by several authors including ourself~\cite{mf}.  Beyond the criticism of the hypotheses, the challenge is to take advantage of the supposed flaw, namely, the lack of accounting for contextuality, to exhibit a classical system that violates Bell's inequalities as well\footnote{Throughout this paper, we use the word `contextual' in its classical meaning, synonymous of `dependent upon the settings'.}. By contrast, the standard wisdom is that `there is no way to form a classical, deterministic, local theory that reproduces quantum theory'~\cite{GHZ}.  However, this impossible challenge has been taken up by several authors, using clever networked computers~\cite{cameleon, hess1} or physical systems~\cite{clover2, matzkin}. So far, these results remain widely controversial~\cite{gill2,gill1,mermin1,nijhoff}. Nevertheless, the present letter falls within the scope of these attempts.

Our proposal, namely a very simple classical dice game, is described in the next section: Two completely separated players select freely their own setting. To insert contextuality, we make use of three different loaded dice: Depending upon the settings, only one die is tossed. In the following  section, we proved that this game meets all features of the EPR experiment. In the subsequent section, owing to the controversial implications of these results, we discuss of some potential objections and point out also a surprising possible spin-off. Finally, we conclude in the last section. 

\section{Description}
\label{description}
We propose to implement a classical game with two remote players \emph{Alice} and \emph{Bob}, one independent referee ${\cal I}$  and a number of three dice: The dice are labelled $\Delta_k$ ($k=1, 2$ and $3$) and the six sides are labelled $\lambda_j$ ($j=1$ to $6$).
The dice are biased: The probability to get the side $\lambda_j$ when rolling the dice $\Delta_k$ is $p_{kj}$, generally different from $1/6$. The probabilities $p_{kj}$ are given in \Tref{dicegame}a. In addition, a binary coefficient $x_{kj}$, given in \Tref{dicegame}b, is attached to each index pair $(k,j)$.  
The two players Alice and Bob located respectively in distant regions ${\cal R}_a$ and ${\cal R}_b$ are only in communication with the remote independent referee ${\cal I}$,  who will throw the dice (\Fref{schema}). 

\newsavebox{\Alice}
\savebox{\Alice}{
{ \xygraph{
!{<0mm,0mm>;<1mm,0mm>:<0mm,1mm>::}
!{(0,0) }*+{}="a"
!{(0,5) }*+{}="b"
!{(30,0) }*+{}="c"
!{(30,5)}*+{}="d"
!{(-7,2.5) }*+{\txt{\fbox{Alice}}}
"c":@/^/"a"^{\txt{then downloads $s_a$}} 
"b":@/^/"d"^{\txt{uploads $k_a$}}
} } 
}
\newsavebox{\Bob}
\savebox{\Bob}{
{ \xygraph{
!{<0mm,0mm>;<1mm,0mm>:<0mm,1mm>::}
!{(0,0) }*+{}="a"
!{(0,5) }*+{}="b"
!{(30,0) }*+{}="c"
!{(30,5)}*+{}="d"
!{(37,2.5) }*+{\txt{\fbox{Bob}}}
"a":@/_/"c"_{\txt{then downloads $s_b$}} 
"d":@/_/"b"_{\txt{uploads $k_b$}}
} }
}
\newsavebox{\Referee}
\savebox{\Referee}{
\fbox{
\xy
(0,20)*++
{\xy
{(0,0)*{}; (8,0)*{};(8,8)*{};(0,8)*{};}
**\frm<1mm>{-};
(4,4)*{\bullet};
\endxy}="x";
(0,10)*++
{\xy
{(0,0)*{}; (8,0)*{};(8,8)*{};(0,8)*{};}
**\frm<1mm>{-};
(2.5,2.5)*{\bullet};(5.5,5.5)*{\bullet};
\endxy}="y";
(0,0)*++
{\xy
{(0,0)*{}; (8,0)*{};(8,8)*{};(8,8)*{};}
**\frm<1mm>{-};
(2.5,2.5)*{\bullet};(4,4)*{\bullet};(5.5,5.5)*{\bullet};
\endxy}="y";
(0,-7)*{\txt{3 dice $\Delta_k$}};
(0,27)*{\txt{Referee ${\cal P}$ }};
\endxy
}
}

\begin{figure}[htb]
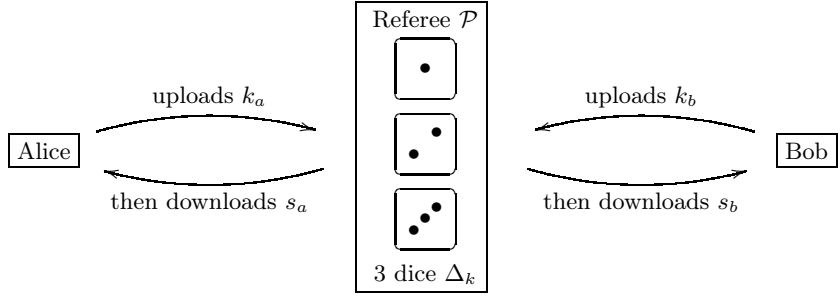
 
\hspace{2cm}
\raisebox{7mm}{\usebox{\Alice}}
\usebox{\Referee}
\raisebox{7mm}{\usebox{\Bob}}
\caption{%
\label{schema} 
{\footnotesize%
 The game is played between two remote parties {Alice} and {Bob}. The goal is to draw two sequences $s_a$ and $s_b\in\pm 1$ violating Bell's inequality. It needs one distant independent referee ${\cal I}$ and a number of three loaded dice $\Delta_k$ labelled $k$ from $1$ to $3$. At each run, each player selects freely one die and submits her/his choice, $k_a$ or $k_b$, to the referee. The referee rolls only one die at random, derives the outcomes $s_a$ and $s_b$ according to the rule of the game given in Table~\ref{dicegame} and sends the results to Alice and Bob
}
} 
\end{figure}

At each run, each player selects freely one die, respectively $\Delta_{k_a}$ and $\Delta_{k_b}$, and submits her/his choice, $k_a$ or $k_b$, to the referee ${\cal I}$. The referee keeps the first received label, or selects one of the two labels at random and ignores the second. For reasons explained later, we will call \emph{gauge selection} this referee choice. Then the referee rolls this only \emph{gauge dice} $\Delta_{k_a}$ or $\Delta_{k_b}$. The outcome of this single trial is a side $\lambda_j$ (not to be mistaken for the final outcome $\pm 1$). The referee will now use the three indices $j,k_a$ and $k_b$ to derive the final results. Let $x_{k_aj}$ and $x_{k_bj}$ be the binary coefficients taken from Table~\ref{dicegame}b. ${\cal I}$ computes $s_a= 2x_{k_aj}-1$ and $s_b=1-2x_{k_bj}$ and transmits these results to Alice and Bob respectively. The final outcome of one run is then the pair $(s_a,s_b)$.

To sum up, each run starts by a pair of free choices $(k_a,k_b)$ and ends by a pair of final outcomes $(s_a,s_b)$. For an ensemble of runs, we claim that the sequences $\{s_a\}$ and $\{s_b\}$ violate the Bell's inequalities.
%
\begin{table} [htb] 
\begin{center}
\subtable 
[
Side probabilities $p_{kj}$
]
{
\begin{tabular}{CCCC}
\br
 \lambda_j &  p_{1j} &  p_{2j}  &  p_{3j}  \\
 \sr
 \lambda_1  &  3/12  &  4/12  &  3/12  \\
 \lambda_2  &  1/12  &  1/12  &  2/12  \\
 \lambda_3  &  2/12  &  1/12  &  1/12  \\
 \lambda_4  &  2/12  &  1/12  &  1/12  \\
 \lambda_5  &  1/12  &  1/12  &  2/12  \\
 \lambda_6  &  3/12  &  4/12  &  3/12  \\
\br
\end{tabular}
}
\qquad
\qquad
\qquad
\subtable 
[
{\footnotesize%
Binary coefficients $x_{kj}$
}
]
{
\begin{tabular}{CCCC}
\br
j &  x_{1j}  &  x_{2j}  &  x_{3j}  \\
 \sr
 1  &  0  &  0  &  0  \\
 2  &  1  &  0  &  0  \\
 3  &  1  &  1  &  0  \\
 4  &  0  &  0  &  1  \\
 5  &  0  &  1  &  1  \\
 6  &  1  &  1  &  1  \\
\br
\end{tabular}
} 
\caption
{%
\footnotesize{%
\label{dicegame} %
Derivation of the outcomes $s_a$ and $s_b$: The three dice are labelled $\Delta_k$ ($k=1, 2$ or $3$) and the six sides are labelled $\lambda_j \ (j=1$ to $6)$. The dice are loaded: The probability to draw the side $\lambda_j$ when rolling the dice labelled $\Delta_k$ is $p_{kj}$. The side probabilities $p_{kj}$ are given in (a). Alice and Bob submit freely their own label to the referee ${\cal I}$, respectively ${k_a}$ and ${k_b}$. The referee rolls at random a single die, either $\Delta_{k_a}$  {or}  $\Delta_{k_b}$,  draws a side $\lambda_j$ and next transmits $s_a=2x_{k_aj}-1$ to Alice and $s_b=1-2x_{k_bj}$ to Bob using the constant binary coefficients $x_{kj}$ given in (b).
}
}
\end{center}
\end{table} 

\section{Analysis}
\label{analysis}

We will first derive a fundamental but surprising result: \emph{The choice of the referee, $\Delta_{k_a}$ or $\Delta_{k_b}$, does not affect the probability system}. Therefore, irrespective of the referee choice, we can compute the conditional probabilities of the final outcomes $s_a$ and $s_b$ given $k_a$ and $k_b$. Furthermore, at each run, Alice and Bob have no way to determine which die, $\Delta_{k_a}$ or $\Delta_{k_b}$, was tossed: {This justifies the term \emph{gauge dice}.}
 
In the following paragraphs, we will show that the probability system meets two basic properties of the EPR experiments, namely, \emph{total correlation} (i.e., if Alice and Bob select the same dice, then they get opposite final outcomes $\pm 1$) and \emph{local consistency} (i.e., in each region, the player is not aware of what happens in the second region and observes that the final outcomes $+1$ and $-1$ are equally likely). 

Then, for an ensemble of runs, when comparing \emph{afterwards} the strings $s_a$ and $s_b$, we will show that the players will record a \emph{violation of Bell's inequalities}. Consequently, we will call this game the \emph{EPR dice game}. 

Finally, we will comment the concept of \emph{instantaneous effect at a distance} in this classical environment.

\subsection {Conditional probabilities and gauge invariance}
\label{proba}
We are now going to compute the conditional probability $\PP(s_a,s_b|k_a,k_b)$ of $(s_a,s_b)$ given $(k_a,k_b)$. This formulation supposes that this probability is uniquely defined, irrespective of the referee choice. We claim that this fundamental condition is satisfied. This may be proved analytically but it is more cogent to check all $36$ possible configurations by hand or by computer program.%
\footnote{%
Actually, we have proceeded in the reverse order, starting from Table~\ref{rho} to derive Table~\ref{dicegame}a, as explained in the last section.
}
For each configuration $(s_a,s_b,k_a,k_b$) we have to consider two cases, $k=k_a$ and $k=k_b$ according to the choice of ${\cal I}$ for the gauge dice $\Delta_k$. Then we collect the relevant sides labelled $\lambda_j$ and we add the probabilities $p_{kj}$.

\begin{table}
\caption{\label {calcul} {\footnotesize Checking gauge invariance: For each configuration $(s_a,s_b,k_a,k_b$) we select the set of relevant sides $\{j\}$ from $(x_{k_aj},x_{k_bj})=[(1+s_a)/2),(1-s_b)/2]$ and Table~\ref{dicegame}b. Then we compare two gauge choices, $k=k_a$ and $k=k_b$ by adding the probabilities $p_{k_aj}$ or $p_{k_bj}$ for the relevant sides in each case. Whatever the choice, we obtain the \emph{same} overall probability $\PP=\PP(s_a,s_b|k_a,k_b)$. See text for the detailed calculation of the underscored configuration.}}
\center{
{\footnotesize
\begin{tabular}{cccccccc}
\br
$s_a$&$s_b$&$k_a$&$k_b$&$\{j\}$&$\sum p_{k_aj}$  & $\sum p_{k_bj}$ & $\PP$ \\
   &   &   &   &     &     $\times 12$ &    $ \times 12 $  &  $ \times 12 $      \\
 \sr
$-1$ &$-1 $& $1$ & $1$  &$-$&$-$&$-$& $0$  \\
$-1$ &$-1 $& $1$ & $2$  &$ \{5\} $    & $1$ & $1$ & $1$  \\
$-1$ &$-1 $& $1$ & $3$  &$ \{4 , 5\} $& $2 + 1$ & $1 + 2$ & $3$  \\
$-1$ &$-1$ & $2$ & $1$  &$ \{2\}  $   & $1$ & $1$ & $1$  \\
$-1$ &$-1$ &$ 2$ & $2$  &$-$&$-$&$-$& $0$  \\
$-1$ &$-1$ & $2$ & $3$  &$ \{4\}     $& $1$ & $1$ & $1$  \\
$-1$ &$-1$ & $3$ & $1$  &$ \{2,  3\}  $& $2 + 1$ & $1 + 2$ & $3$ \\
$-1$ &$-1 $& $3$ & $2$  &$ \{3\} $&$ 1 $&$ 1 $&$ 1 $ \\
$-1$ &$-1$ & $3$ & $3$  &$-$&$-$&$-$& $0$  \\
$-1$ & $+1$ & $1$ & $1$ &$ \{1,  4,  5\} $&$ 3 + 2 + 1 $&$ 3 + 2 + 1 $&$ 6 $ \\
\hline
$-1$ & $+1$ & $1$ & $2$ &$ \{1,  4\} $&$ 3 + 2 $&$ 4 + 1 $&$ 5 $ \\
\hline
$-1$ & $+1$ & $1$ & $3$ &$ \{1\} $& $3$ & $3$ & $3$  \\
$-1$ & $+1$ & $2$ & $1$ &$ \{1,  4\} $&$ 4 + 1 $&$ 3 + 2 $&$ 5$  \\
$-1$ & $+1$ & $2$ & $2$ &$ \{1,  2,  4 \} $&$ 4 + 1 + 1 $&$ 4 + 1 + 1 $& $6$  \\
$-1$ & $+1$ & $2$ & $3$ &$ \{1,  2\} $&$ 4 + 1 $& $3 + 2 $& $5$  \\
$-1$ & $+1$ & $3$ & $1$ &$ \{1\} $& $3$ & $3$ & $3$  \\
$-1$ & $+1$ & $3$ & $2$ &$ \{1,  2\} $&$ 3 + 2 $& $4 + 1$ & $5$  \\
$-1$ & $+1$ & $3$ & $3$ &$ \{1,  2,  3\} $& $3 + 2 + 1$ & $3 + 2 + 1$ & $6$  \\
 $+1$ &$-1$ & $1$ & $1$ &$ \{2,  3,  6\} $&$ 1 + 2 + 3 $&$ 1 + 2 + 3 $&$ 6$  \\
$ +1$ &$-1$ & $1$ & $2$ &$ \{3,  6\} $&$ 2 + 3 $&$ 1 + 4 $&$ 5$  \\
$ +1$ &$-1$ & $1$ & $3$ &$ \{6\} $&$ 3 $&$ 3 $& $3$  \\
$ +1$ &$-1$ & $2$ & $1$ &$ \{3,  6\} $& $1$  +$4$ & $2$ + $3$ & $5$  \\
 $+1$ &$-1$ & $2$ & $2$ &$ \{3,  5,  6\} $&$ 1 + 1 + 4 $&$ 1 + 1 + 4 $&$ 6$  \\
$ +1$ &$-1$ & $2$ & $3$ &$ \{5,  6\} $& $1 + 4 $& $2 + 3$ & $5$  \\
$ +1$ &$-1$ & $3$ & $1$ &$ \{6\} $& $3$ & $3$ & $3$  \\
$ +1 $&$-1$ & $3$ & $2$ &$ \{5,  6\} $& $2 + 3$ & $1 + 4$ & $5$  \\
$ +1 $&$-1$ & $3$ & $3$ &$ \{4,  5,  6\} $&$ 1 + 2 + 3 $&$ 1 + 2 + 3 $& $6$  \\
$ +1$ & $+1$ & $1$ & $1$ &$-$&$-$&$-$& $0$  \\
$ +1 $&$ +1$ & $1$ & $2$ &$ \{2\} $& $1$ & $1$ & $1$  \\
$ +1 $&$ +1 $& $1$ & $3$ &$ \{2,  3\} $&$ 1 + 2 $&$ 2 + 1 $&$ 3 $ \\
$ +1 $&$ +1 $&$ 2 $&$ 1 $&$ \{5\} $&$ 1 $&$ 1 $&$ 1$  \\
$+1$ & $+1$ & $2$ & $2$&$-$&$-$&$-$& $0$  \\
$ +1 $&$ +1 $&$ 2 $&$ 3 $&$ \{3\} $&$ 1 $&$ 1 $&$ 1 $ \\
$ +1 $&$ +1 $&$ 3 $&$ 1 $&$ \{4,  5\} $&$ 1 + 2 $&$ 2 + 1 $&$ 3 $ \\
$ +1 $&$ +1 $&$ 3 $&$ 2 $&$ \{4\} $&$ 1 $&$ 1 $&$ 1  $\\
$ +1 $&$ +1 $&$ 3 $&$ 3 $&$-$&$-$&$-$&$ 0  $\\
\br
\end{tabular}
}
}
\end{table}

Suppose for example that Alice submits the die $\Delta_1$ ($k_a=1$) and Bob the die $\Delta_2$ ($k_b=2$) (Table~\ref{calcul}). To compute $\PP(s_a,s_b|1,2)$, we have to consider two cases, according to the choice of ${\cal I}$. Assume that $s_a=-1$ and $s_b=+1$, i.e., $(x_{1j},x_{2j})= [(1+s_a)/2),(1-s_b)/2]=(0,0)$. From inspection of columns $x_{1j}$ and $x_{2j}$ of Table~\ref{dicegame}b, we find two solutions $j= 1$ and $j=4$.  

Case \#1 : $k=1$ (${\cal I}$  selects the dice $\Delta_1$). Then from rows $\lambda_1$ and $\lambda_4$ and column $p_{1j}$ of Table~\ref{dicegame}a, we obtain $\PP(-1,+1|1,2)=p_{11}+p_{14}=3/12 +2/12=5/12.$ 

Case \#2 : $k=2$ (${\cal I}$  selects the dice $\Delta_2$). Then from rows $\lambda_1$ and $\lambda_4$ and column $p_{2j}$ of Table~\ref{dicegame}a, we obtain $\PP(-1,+1|1,2)=p_{21}+p_{24}=4/12 +1/12=5/12.$

The two values are identical. We proceed similarly for all possible configurations. The details of the computation are given in Table~\ref{calcul}. For each configuration, we find that the probability $\PP(s_a,s_b|k_a,k_b)$ does not depend upon the choice of the referee. We will call  this property \emph{gauge invariance}.

\emph{This gauge invariance is fundamental : the particular die tossed by the referee can be \emph{any} of the two dice chosen by Alice and Bob. This is the core of the EPR paradox.} 

The final results are given in Table~\ref{rho}. 


\begin{table}
\caption{\label {rho} {\footnotesize Conditional probability $\PP(s_a,s_b|k_a,k_b)$ to get $s_a$ and $s_b$ when Alice chooses $k_a$ and Bob $k_b$. The fundamental point is that $\PP(s_a,s_b|k_a,k_b)$ does not depend on the choice of the referee. It is easy to verify that Bell's inequalities, \Eref{bellineq} are violated.}}
\center{
\begin{tabular}{cccc}
\br
&                     & (1,2) & \\
$\quad \ (k_a, k_b)\rightarrow$ & (1,1) & (2,1) & \\
$(s_a, s_b)$& (2,2) & (2,3) &(1,3) \\
$\downarrow$ & (3,3) & (3,2) &(3,1) \\
 \sr
$(-1, +1)$ & $1/2$ & $5/12$ & $1/4$\\
$(-1, -1)$ & $0$ & $1/12$ & $1/4$ \\
$(+1, +1)$ & $0$ & $1/12$ & $1/4$ \\
$(+1, -1)$ & $1/2$ & $5/12$ & $1/4$\\
\br
\end{tabular}
}
\end{table}


\subsection{Total correlation}
\label{correlation}
From Table~\ref{rho}, it is clear that if Alice and Bob choose the same dice, $\Delta_a = \Delta_b$, then $s_a=-s_b$.
We will call this property \emph{total correlation}. 

\subsection{Local consistency}
\label{local}
In region $ {\cal R}_b$, when Bob selects his own dice, he receives, after a delay, his final outcome $s_b=\pm 1$. This means that the referee has selected a gauge dice and has performed a trial. But this is not observable from ${\cal R}_a$ and Alice is not aware of what happens in ${\cal R}_b$. She independently selects her own dice and next downloads her final outcome $s_a$. From Table~\ref{rho} it is easily seen by inspection that the marginal probability $\pr(s_a|k_a)$ of $s_a$ given $k_a$, does not depend on $k_b$. Whatever $k_b$, we have
\begin{equation}
\label{locality}
\pr(s_a|k_a) = \PP(s_a,1|k_a,k_b) + \PP(s_a,-1|k_a,k_b)=1/2
\end{equation} 
In other words, irrespective of $k_b$, the two final outcomes $s_a=\pm 1$ are equally likely. This fundamental property characterizes quantum Bell-type experiments, and is compatible with space-like separation of the two regions. We will call this property \emph{local consistency}.

\subsection{Violation of Bell's inequalities}
\label{bell}
From Table~\ref{rho}, even when the dice selected by Alice and Bob are different, the two final outcomes are correlated. We are going to show that this correlation can lead to a violation of Bell's inequalities. 

There is number of formulations of Bell's inequalities. For clarity, let us select the initial formulation by Bell, that reads
\begin{equation}
\label{bellineq}
| \s(k_1,k_2)- \s(k_1,k_3)| \le 1+ \s(k_2,k_3),
\end{equation}
where $ \s(k_a,k_b)=\EE[s_a  s_b]$ is the expectation value of $s_a s_b$ when Alice and Bob select respectively the dice labelled $k_a$ and $k_b$.
{In his famous theorem, Bell assumes the existence of an absolute probability system irrespective of the particular settings.  By contrast, accounting for contextuality, we have to compute the expectation value with respect to the conditional probability $\PP(s_a,s_b|k_a,k_b)$.}

Suppose that $k_1=1$, $k_2=2$ and $k_3 =3$. From Table~\ref{rho} we compute easily,
$ \s(1,2)=  \s(2,3) = (5/12)(-1)(+1) + (1/12)(-1)(-1)+(1/12)(+1)(+1)+(5/12)(+1)(-1)=-2/3$, and 
$ \s(1,3)= (1/4)(-1)(+1) + (1/4)(-1)(-1)+(1/4)(+1)(+1)+(1/4)(+1)(-1)=0.$ Then
$$| \s(1,2)- \s(1,3)| =2/3 > 1+ \s(2,3) = 1/3.$$
Bell's inequality, \Eref{bellineq}, is violated.

Another popular formulation, involving four settings, $k_{a}, k_{a'}, k_{b}$ and $ k_{b'}$, is the CHSH inequality~\cite{chsh}, that reads 
\begin{equation}
\label{CHSH}
|  \s( k_{a}, k_{b}) +  \s(k_{a}, k_{b'}) +  \s(k_{a'}, k_{b}) -  \s(k_{a'}, k_{b'}) | \le 2.
\end{equation}

Alice has to choose between two settings, $k_{a}$ and $k_{a'}$, while Bob  chooses from $k_{b}$ and $k_{b'}$. 
This equation makes use of 4 settings while we have only 3 dice at disposal. Thus, one die will be used twice. Select for instance $k_{a} =1$, $k_{a'}=3$, $k_{b}=2$ and $ k_{b'}=1$.
We have to compute $\s(1,1) =-1/2-1/2 =-1$. 
Thus,

$$ | \s(1,2 ) +  \s(1,1) +  \s(3,2) -  \s(3,1)| = |-{2}/{3}-1-{2}/{3}-0| = 2+{1}/{3} > 2.$$
Again, CHSH-inequality Eq.(\ref{CHSH}) is violated.

\subsection{Instantaneous effect at a distance}
\label{instant}
The surprising apparent instantaneous effect at a distance of quantum mechanics was pictured by Jaynes~\cite{clearing} as follows: `The spooky supraluminal stuff [\dots] disappears as soon as we recognize [\dots] that what is travelling faster than light is not a physical causal influence, but only a logical inference.' To clarify concretely this point, suppose that a demon rolls a pair of dice in a distant planet around Betelgeuse and that the outcome is a double-six. This nice result is immediately \emph{true} on the Earth. Nevertheless, in accordance with Lorentz covariance, we will have to wait for at least 427 years before we could learn this good news. Therefore, on the one hand, we may consider that the instantaneous event is purely fictitious on the Earth. But on the other hand, the same instantaneous event may be considered as real since \emph{afterwards}, we will be able to derive exactly its date and its location. In other words, the score of the demon may be considered as instantaneously valid at a distance. Let us call this trivial paradox \emph{Betelgeuse effect}. 

The Betelgeuse effect holds in the EPR dice game. Due to gauge invariance, as soon as the first player, Alice or Bob, has selected her or his own dice, this first die becomes the gauge dice and the probability system (\Tref{rho}) is completed. The first final outcome becomes available. Nevertheless, we have to wait for the second selection before learning the second final outcome. But \emph{afterwards}, when analysing the record of a number of runs, the relevant time to account for remains the instant where the probability system was completed. This can lead to apparent instantaneous effects at a distance or even to apparent violation of causality.

For example, suppose that Bob submits first his dice and that Alice waits a good while. Later, as soon as Alice chooses her own dice, her choice and her final outcome are valid instantaneously at a distance in the Bob region. This will be recorded afterwards, when Alice and Bob will observe the correlation of their outcomes. 

In other words, the EPR dice game exhibits both a form of apparent instantaneous effect at a distance and an apparent causality paradox similar to the Wheeler's delayed choice gedanken experiment~\cite{wheeler}.

\section{Discussion}
\label{discussion}
Since the above results are clearly conflicting with the standard wisdom, they deserve careful thought. In this section, we discuss of the main potential objections.%
\footnote{%
We have to thank a number of anonymous referees from six different journals who raised these objections (and for good measure, recommanded the rejection of this paper). The present formulation use often their own wordings.
}
 We also point out a possible spin-off.

\subsection{Analogy with a physical experiment}

The first question is the relevance of the model to the experimental set-up envisaged by Bell. Of course,  we do not expect that the present dice game describes a real physical situation but, in the present game the concept of particle vanishes.  Thus, at first glance, when examining \Fref{schema}, one might suspect that the referee acts genuinely as the source of particles and therefore that the settings have to be selected before the photons are emitted. Actually, the model does not in the least require such a time inversion and the source is not allowed to access the measurement settings. This can be proved by the following scheme, where initially, a source (not represented on \Fref{schema}) launches the photons and subsequently, the referee triggers the quantum mechanical collapse~\cite{mf}: 

At the beginning of each run, a source launches a pair of entangled photons towards two distant observers Alice and Bob. This is the only role of the source. Next, the observers select independently their polarization. Later, when they receive their own photon, they inform a distant ignition point ${\cal I}$, independent of the source, on which polarization they decided to make the measurement. Basing on a specific random process,  ${\cal I}$ (disguised as the referee) triggers the collapse as soon as he receives the first information. Then he sends the results to the observers, so that afterwards, strange correlations violating Bell inequalities are observed. 

For example,  a similar mechanism, using time-retarded field, has been proposed by Clover~\cite{clover2} to explain the Innsbruck experiment~\cite{innsbruck}. 
Since this is a different and complex story, we will not elaborate further, but, where we are concerned, \emph{this scheme secures the compatibility of the analogy with a standard \emph{EPR} experiment.}

\subsection{Local realism}
We now address the question of local realism. We must emphasize that the above dice game is strictly classical. It is not a thought experiment, because it can be trivially implemented in the every day classical world. It should not be mistaken for a `non local game' as defined by Cleve et al~\cite{cleve,silman} nor for any kind of `teleportation'~\cite{bennett}, which makes use of quantum components \emph{together} with `local operations and classical communication' to reconstruct quantum correlations. Classical physics is clearly the bench mark of local realism. Therefore, \emph{local realism is secured by plain classicality.}

The first element of local realism is realism itself. However, one might consider that the referee does not roll the dice for both Alice and Bob but only for one of them, say Alice. From gauge invariance the joint probability of the two outcomes is nevertheless completed.  Bob, whose dice may not have been rolled, is told the outcome of the unperformed measurement `on his own dice' and this could be viewed as `counterfactual'. But actually, this `paradox' is quite artificial, since Bob dice, according to the rules of the game, simply need not be rolled. Furthermore, there are six gauge outcomes (from $1$ to $6$) but just two final outcomes ($\pm 1$) and, out of the referee black box, the only variables are the settings $k_a, k_b$  and the final outcomes $s_a, s_b$. Neither Alice nor Bob are concerned by the dice. We have introduced a triplet of dice only to illustrate a possible (but not necessary) mechanism of gauge probability. Therefore, quite the opposite, this situation points out the very core of Bell's theorem flaw: A simple toss, is unable to describe a contextual phenomenon, be it classical or quantum. Thus, more sophisticated tools are required, such as the present stochastic gauge system. In this model the eventual expectation of Bob to obtain the outcome of `his own dice' is simply irrelevant and does not affect the  property of \emph{realism}.

The second element of local realism is locality, which deserves further discussion.

\subsection{Space-like separation}

Local realism does not exclude classical communication but only supraluminal communication. In the present model, all information is transmitted with finite velocity: Clearly, locality holds. 
Nevertheless, if time is not an issue in the proof of Bell's theorem, it is crucial to Einstein-Podolsky-Rosen's argument that the two observers should not communicate. 

In the present dice game, Alice and Bell are not allowed to directly communicate,  but both are in communication with the referee. One might consider that this is an indirect linkage. It is fundamental to clarify this point because if there were possible transmission of information between the two players, the two regions could not be considered as space-like separated. Actually, owing to local consistency, \Eref{locality},  \emph{no information} can be transmitted between Alice and Bob~\cite{mf}. For example, suppose that Alice selects first her dice $k_a$. Bob has no way to guess her choice. When he selects independently his own dice, he has no extra resource at his disposal. By contrast, the problem of classically simulating quantum correlations~\cite{maudlin,brassard,steiner,degorre1,regev} makes use of a  classical transmission channel to transmit data-sets \emph{between Alice and Bob}. In the present game, no such resource exists. \emph{Space-like separation is secured by the fact that no information can be transmitted between Alice and Bob.}

On the other hand, the link between each player and the referee may have a deep significance as we are going to elaborate in the next paragraph.

\subsection{Bell's inequality violation and the holographic hypothesis} 
\label{holography}
By definition, the referee $\cal I$, i.e., the stochastic system, is located at the boundary of the two regions and we have shown that \emph{no information can cross this frontier}. Therefore, in each region, a certain amount of information is hidden, which corresponds to a certain amount of entropy located in the random system. Due to symmetry, this `gauge entropy' is likely to be shared by the two parties and could be related with the information required to classically simulate quantum correlation~\cite{brassard}. This situation is arguably encountered in quantum mechanics: Consider a pair of entangled entities located in two space-like separated regions. By definition, the boundary between the two regions is a light cone, or in other words, a causal horizon~\cite{jacobson}. This horizon is the only geometrical feature shared by the two parties. This argument has been used by Srednicki~\cite{srednicki} to justify that the entanglement entropy of a scalar field has to be proportional to the area of the horizon in the context of black hole physics. This was later the base of the holographic hypothesis~\cite{bousso} of 't Hooft~\cite{hooft} and Susskind~\cite{susskind}. In the present dice game, Alice and Bob share a common amount of entropy similarly located at the boundary between the two regions. Thus, this `gauge entropy' imposed by contextuality, i.e., Bell's inequality violation, could be a structural feature of entangled entities separated by a horizon. This suggests a link between Bell's inequality violation and the holographic hypothesis. 

\subsection {Hints on the design of classical contextual models}
\label{models}
For clarity, we have chosen to detail the simplest classical device exhibiting a violation of Bell's inequalities, \Eref{bellineq}, with just three settings and two final outcomes $\pm 1$. However, more complex systems with any number of settings can be similarly constructed. The discussion gets off the scope of this paper but we are going to give some hints. To construct a discrete system with $K$ settings one has to choose a particular target similar to \Tref{rho} and derive $K$ probability distributions with a number of $2^K$ possible outcomes for each distribution. It is convenient to define a set of binary coefficients $x_{kj}$ (similar to one column labelled $k$ of \Tref{dicegame}b) as the $k^{\ordinal}$ binary digit in the binary expansion of $j-1$ for each outcome labelled $j$. The solution (similar to one column of \Tref{dicegame}a) is given by a degenerate linear system with $2K$ independent equations for $2^K$ unknowns. This allows to decrease to $2K$ the number of actual outcomes (e.g., $6$ outcomes in the present dice game, or $8$ outcomes for a $4$-setting device). In addition, we have previously described a model with a continuous ensemble of settings, which exactly simulates the EPR-B experiment~\cite{mf}. More complex contextual models with more than two final outcomes and more than two regions are likely to be constructed similarly.

\section{Conclusion}
\label{conclusion}

Our goal was to construct a classical system meeting all the features of the EPR-B experiment. Firstly, we have taken advantage of a well known flow in the assumptions sustaining the derivation of Bell's theorem. The second ingredient was the use of a new concept~\cite{mf}, that we have called \emph{stochastic gauge system}, which allows the definition of a convenient contextual probability space.  We have then designed a very simple dice game between two parties, Alice and Bob who cannot exchange information and consequently can be regarded as space-like separated. Nevertheless, they record a violation of Bell's inequalities. Therefore, we claim that we have succeeded in our initial goal: This proves that a strictly classical system can violate Bell's inequality. In the language of Bell~\cite{bell1}, plain classicality implies `local realism': `Any correlation between measurements performed at different places should derive from events which happened in the intersection of the past light cones of the measurements.' The assumption that Bell's theorem provides necessary conditions for `local realism' is disproved. This supports a speculation by Santos~\cite{santos},  who suspects that ``the `resistance' of detection loophole to be eliminated could point out a practical impossibility of falsifying local realism''. This model respects the principle of Lorentz covariance and provides a straightforward solution to several quantum mechanical paradoxes, even if only experiment will finally decide.

\bibliography{biblio}

\begin{thebibliography}{10}
\expandafter\ifx\csname url\endcsname\relax
  \def\url#1{\texttt{#1}}\fi
\expandafter\ifx\csname urlprefix\endcsname\relax\def\urlprefix{URL }\fi
\expandafter\ifx\csname href\endcsname\relax
  \def\href#1#2{#2} \def\path#1{#1}\fi

\bibitem{bell}
J.~S. Bell, On the {E}instein-{P}odolsky-{R}osen paradox, Physics 195 (1964) 1.

\bibitem{bell1}
J.~S. Bell, Speakable and unspeakable in quantum mechanics, Cambridge
  University Press, Cambridge, 1987.

\bibitem{einstein}
A.~Einstein, B.~Podolsky, N.~Rosen, Can quantum mechanical description of
  reality be considered complete?, Phys. Rev. 47 (1935) 777.

\bibitem{aspect}
A.~Aspect, J.~Dalibard, G.~Roger, Experimental test of {B}ell's inequalities
  using time-varying analysers, Phys. Rev. Lett. 49 (1982) 1804--1807.

\bibitem{genovese}
M.~Genovese, Research on hidden variable theory: a review of recent progresses,
  Physics Reports 413 (2005) 319, (and reference therein).
\newblock \href {http://arxiv.org/abs/quant-ph/0701071v1}
  {\path{arXiv:quant-ph/0701071v1}}.

\bibitem{GHZ}
D.~M. Greenberger, M.~A. Horne, A.~Zeilinger, Going beyond {B}ell's theorem,
  in: M.~Kafatos (Ed.), Bell's Theorem, Quantum Theory, and Conceptions of the
  Universe, Kluwer, Dordrecht, 1989, pp. 69--72.
\newblock \href {http://arxiv.org/abs/0712.0921v1 [quant-ph]}
  {\path{arXiv:0712.0921v1 [quant-ph]}}.

\bibitem{santos}
E.~Santos, The failure to perform a loophole-free test of {B}ell’s inequality
  supports local realism, Foundations of Physics 34~(11) (2004) 1643--1673.
\newblock \href {http://arxiv.org/abs/quant-ph/0410193}
  {\path{arXiv:quant-ph/0410193}}.

\bibitem{clearing}
E.~T. Jaynes, \href{http://bayes.wustl.edu/etj/articles/cmystery.pdf}{Clearing
  up mysteries}, in: J.~Skilling (Ed.), Maximum entropy and {B}aysian methods,
  Kluwer, Dordrecht, 1989, pp. 1--29.
\newline\urlprefix\url{http://bayes.wustl.edu/etj/articles/cmystery.pdf}

\bibitem{mf}
M.~Feldmann, New loophole for the {E}instein-{P}odolsky-{R}ozen paradox, Found.
  Phys. Lett. 8~(1) (1995) 41--53.
\newblock \href {http://arxiv.org/abs/quant-ph/9904051}
  {\path{arXiv:quant-ph/9904051}}, \href {http://dx.doi.org/10.1007/BF02187530}
  {\path{doi:10.1007/BF02187530}}.

\bibitem{kracklauer}
A.~F. Kracklauer, The error in {B}ell's theorem (1998).
\newblock \href {http://arxiv.org/abs/quant-ph/9810081}
  {\path{arXiv:quant-ph/9810081}}.

\bibitem{hess1}
K.~Hess, W.~Philipp, Einstein-separability, time related hidden parameters for
  correlated spins, and the theorem of {B}ell, Proc. Nat. Acad. Sciences USA
  98~(7) (2002) 14224, 14228.

\bibitem{khrennikov}
A.~Y. Khrennikov, Interpretations of Probability, VSP Int. Sc. Publishers,
  Utrecht, 1999.

\bibitem{khrennikov2}
A.~Khrennikov, I.~Volovich, Local realism, contextualism and loopholes in
  {B}ell`s experiments (2002).
\newblock \href {http://arxiv.org/abs/quant-ph/0212127}
  {\path{arXiv:quant-ph/0212127}}.

\bibitem{cameleon}
L.~Accardi, Urne e camaleonti. Dialogo sulla realtà, le leggi del caso e la
  teoria quantistica, Il Saggiatore, 1997.

\bibitem{clover1}
M.~Clover, Bell's theorem: {A} critique (2005).
\newblock \href {http://arxiv.org/abs/quant-ph/0502016}
  {\path{arXiv:quant-ph/0502016}}.

\bibitem{nieuwenhuizen}
T.~M. {Nieuwenhuizen}, {Where {B}ell Went Wrong}, in: {L.~Accardi, G.~Adenier,
  C.~Fuchs, G.~Jaeger, A.~Y.~Khrennikov, J.-{\AA}.~Larsson, \& S.~Stenholm}
  (Ed.), American Institute of Physics Conference Series, Vol. 1101, 2009, pp.
  127--133.
\newblock \href {http://arxiv.org/abs/0812.3058} {\path{arXiv:0812.3058}}.

\bibitem{adenier}
G.~Adenier, A refutation of {B}ell's theorem (2000).
\newblock \href {http://arxiv.org/abs/quant-ph/0006014}
  {\path{arXiv:quant-ph/0006014}}.

\bibitem{clover2}
M.~Clover, The {I}nnsbruck {E}{P}{R} experiment: {A} time-retarded local
  description of space-like separated correlations (2003).
\newblock \href {http://arxiv.org/abs/quant-ph/0304115}
  {\path{arXiv:quant-ph/0304115}}.

\bibitem{matzkin}
A.~Matzkin, \href{doi:10.1103/PhysRevA.77.062110}{Bell's theorem as a signature
  of nonlocality: a classical counterexample}, Physical Review A 77 (2008)
  062110.
\newblock \href {http://arxiv.org/abs/0709.2114v2 [quant-ph]}
  {\path{arXiv:0709.2114v2 [quant-ph]}}.
\newline\urlprefix\url{doi:10.1103/PhysRevA.77.062110}

\bibitem{gill2}
R.~D. Gill, Accardi contra {B}ell (cum mundi): The impossible coupling, in:
  M.~Moore, S.~Froda, C.~Léger (Eds.), Mathematical Statistics and
  Applications: {F}estschrift for {C}onstance van {E}eden, Institute of
  Mathematical Statistics, 2003, pp. 133--154.
\newblock \href {http://arxiv.org/abs/quant-ph/0110137}
  {\path{arXiv:quant-ph/0110137}}.

\bibitem{gill1}
R.~D. Gill, G.~Weihs, A.~Zeilinger, M.~Zukowski, No time loophole in {B}ell's
  theorem; the {H}ess-{P}hilipp model is non-local, PNAS 99 (2002) 14632.
\newblock \href {http://arxiv.org/abs/quant-ph/0208187}
  {\path{arXiv:quant-ph/0208187}}.

\bibitem{mermin1}
N.~D. Mermin, Bell's theorem in the presence of classical communication (2002).
\newblock \href {http://arxiv.org/abs/quant-ph/0207140}
  {\path{arXiv:quant-ph/0207140}}.

\bibitem{nijhoff}
R.~Nijhoff, A local hidden-variable model violating {B}ell's inequalities: a
  reply to {M}atzkin (2008).
\newblock \href {http://arxiv.org/abs/0807.1815v1 [quant-ph]}
  {\path{arXiv:0807.1815v1 [quant-ph]}}.

\bibitem{chsh}
J.~F. Clauser, M.~A. Horne, A.~Shimony, R.~A. Holt, Proposed experiment to test
  local hidden-variable theories, Phys. Rev. Lett. 23~(15) (1969) 880--884.
\newblock \href {http://dx.doi.org/10.1103/PhysRevLett.23.880}
  {\path{doi:10.1103/PhysRevLett.23.880}}.

\bibitem{wheeler}
J.~A. Wheeler, W.~H. Zurek, Quantum theory and Measurement, Princeton
  University Press, 1984.

\bibitem{innsbruck}
G.~Weihs, T.~Jennewein, C.~Simon, H.~Weinfurter, A.~Zeilinger, Violation of
  {B}ell's inequality under strict {E}instein locality conditions, Physical
  Review Letters 81 (1998) 5039.
\newblock \href {http://arxiv.org/abs/quant-ph/9810080}
  {\path{arXiv:quant-ph/9810080}}.

\bibitem{cleve}
R.~Cleve, P.~Høyer, B.~Toner, J.~Watrous, Proceedings of the 19th {I}{E}{E}{E}
  conference on computational complexity (2004).

\bibitem{silman}
N.~A. J.~Silman, S.~Machnes, On the relation between {B}ell inequalities and
  nonlocal games (2007).
\newblock \href {http://arxiv.org/abs/0710.3322v2 [quant-ph]}
  {\path{arXiv:0710.3322v2 [quant-ph]}}.

\bibitem{bennett}
C.~H. Bennett, G.~Brassard, C.~Crepeau, R.~Jozsa, A.~Peres, W.~K. Wootters,
  Teleporting an unknown quantum state via dual classical and
  {E}instein-{P}odolsky-{R}osen channels, Phys. Rev. Lett. 70 (1993) 1895.

\bibitem{maudlin}
T.~W. Maudlin, Bell’s inequality, information transmission, and prism models,
  in: D.~Hull, M.~Forbes, K.~Okruhlik (Eds.), Biennal Meeting of the Philosophy
  of Science Association, Philosophy of Science Association, East Lansing,
  1992, pp. 404--417.

\bibitem{brassard}
G.~Brassard, R.~Cleve, A.~Tapp, The cost of exactly simulating quantum
  entanglement with classical communication, Physical Review Letters 83 (1999)
  1874.
\newblock \href {http://arxiv.org/abs/quant-ph/9901035}
  {\path{arXiv:quant-ph/9901035}}.

\bibitem{steiner}
M.~Steiner, Towards quantifying non-local information transfer: Finite-bit
  non-locality, Physics Letters A 270 (2000) 239.
\newblock \href {http://arxiv.org/abs/quant-ph/9902014}
  {\path{arXiv:quant-ph/9902014}}.

\bibitem{degorre1}
J.~Degorre, S.~Laplante, J.~Roland, Simulating quantum correlations as a
  distributed sampling problem, Physical Review A 72 (2005) 062314.
\newblock \href {http://arxiv.org/abs/quant-ph/0507120}
  {\path{arXiv:quant-ph/0507120}}.

\bibitem{regev}
O.~Regev, B.~Toner, Simulating quantum correlations with finite communication,
  focs 0 (2007) 384--394.
\newblock \href {http://arxiv.org/abs/0708.0827[quant-phys]}
  {\path{arXiv:0708.0827[quant-phys]}}, \href {http://dx.doi.org/http:
  //doi.ieeecomputersociety.org/10.1109/FOCS.2007.62} {\path{doi:http:
  //doi.ieeecomputersociety.org/10.1109/FOCS.2007.62}}.

\bibitem{jacobson}
T.~Jacobson, R.~Parentani, Horizon entropy, Found. Phys. 33 (2003) 323.
\newblock \href {http://arxiv.org/abs/gr-qc/0302099}
  {\path{arXiv:gr-qc/0302099}}.

\bibitem{srednicki}
M.~Srednicki, Entropy and area, Physical Review Letters 71 (1993) 666.
\newblock \href {http://arxiv.org/abs/hep-th/9303048}
  {\path{arXiv:hep-th/9303048}}.

\bibitem{bousso}
R.~Bousso, The holographic principle, Reviews of Modern Physics 74 (2002) 825.
\newblock \href {http://arxiv.org/abs/hep-th/0203101}
  {\path{arXiv:hep-th/0203101}}.

\bibitem{hooft}
G.~'t~Hooft, The holographic principle (2000).
\newblock \href {http://arxiv.org/abs/hep-th/0003004}
  {\path{arXiv:hep-th/0003004}}.

\bibitem{susskind}
L.~Susskind, The world as a hologram, J.Math.Phys. 36 (1995) 6377--6396.
\newblock \href {http://arxiv.org/abs/hep-th/9409089v2}
  {\path{arXiv:hep-th/9409089v2}}.

\end{thebibliography}

\end{document}